---

# An Efficient Bit Plane X-ORing Algorithm for Irreversible Image Steganography

---

Soumendu Chakraborty, Anand Singh Jalal* and Charul Bhatnagar

GLA University,
Mathura-281406, India
E-mail: soum.uit@gmail.com
E-mail: anandsinghjalal@gmail.com
E-mail: vrij@iiita.ac.in
*Corresponding author

**Abstract:** The science of hiding secret information in another message is known as Steganography; hence the presence of secret information is concealed. It is the method of hiding cognitive content in same or another media to avoid recognition by the intruders. This paper introduces new method wherein irreversible steganography is used to hide an image in the same medium so that the secret data is masked. The secret image is known as payload and the carrier is known as cover image. X-OR operation is used amongst mid level bit planes of carrier image and high level bit planes of data image to generate new low level bit planes of the stego image. Recovery process includes the X-ORing of low level bit planes and mid level bit planes of the stego image. Based on the result of the recovery, subsequent data image is generated. A RGB color image is used as carrier and the data image is a grayscale image of dimensions less than or equal to the dimensions of the carrier image. The proposed method greatly increases the embedding capacity without significantly decreasing the PSNR value.

## 1 Introduction

Modern day's communication requires high level of security in transmission. There are two ways of achieving this: one by securing the channel and the other is by securing the message. Steganography is a well known and widely used technique that manipulates information (messages) in order to hide their existence. This technique has many applications in computer science and other related fields: It is used to protect military messages, corporate data, personal files, etc.

Steganography is the art and science of communication which hides the presence of information (Clair at al., 2001). It conceals the very existence of the message by engrafting it inside a carrier file of some type. A snooper can intercept an encrypted message;

however he may not even know whether a steganographic message exists. Steganography, attempts to prevent an intruder from suspecting that the data is there (Westfeld et al., 1998).The goal of steganography is to avoid drawing attention to the transmission of the secret message.

On the other hand, steganalysis is a way of detecting possible secret communication using steganography. That is, steganalysis attempts to beat steganography techniques. It relies on the fact that hiding information in digital media amends the carrier and introduces unusual signatures or some form of debasement that could be exploited. Thus, it is crucial that a steganography system make certain that hidden messages are not detectable (Lin et al., 2004; Rabah, 2004; Morkel et al., 2005). Steganography includes the hiding of various media like text, image, audio, video files in another media of same type or of different type, before the message hidden in the selected media is transmitted to recipient. At the receiver's end, reverse procedure is implemented to recover the original information (Krenn, R.).

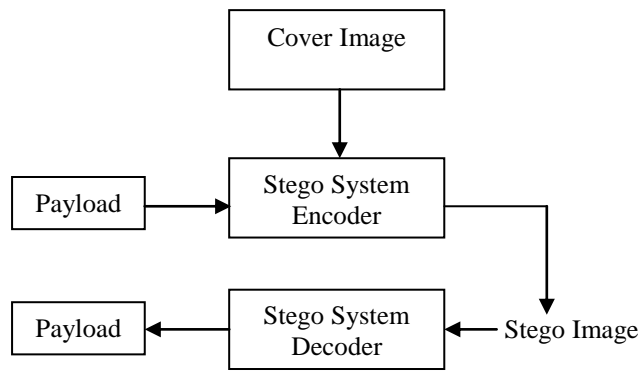

Figure 1: Steganographic Flow

The general process of steganography is shown in figure 1. Two fundamental expectations that a steganography algorithm should live up to are: (1) large number of secret bits can be embedded within the host image, so the algorithm must efficiently implement it, (2) the visual noise introduced due to embedding of the secret data should be at minimum level. Imperceptible stego-image quality is the most important feature of any steganography algorithm. There are several techniques that can be used to hide secret information inside an image like LSB Method, injection, substitution, and generation (Krenn, R.). Most of the existing algorithms do not explore the three dimensional characteristics of a RGB image for data hiding. Our quest is to embed secret data across different dimensions of a RGB image. In the proposed method we have used the concept of injection wherein payload is substituted in parts of the carrier.

## 2 Related work

In the area of hidden information extraction from a stego image intense research work is going on around steganalysis. To get the better of steganalysis schemes many steganography schemes have been proposed, which explore the multiple dimensions of cover image. Many ideas and techniques have been proposed to secure data i.e., mainly concealing of text in images. The simple method to do the same is Least Significant Bit (LSB) replacement method. However, it has its own limitations (Anderson et al., 2001).Steganalysis can be easily done on LSB replacement technique (Dumitrescu et al., 2002).

Cheeldod et al. (2008) proposed an adaptive steganographic approach that selects the specific region of interest (ROI) in the cover image. Data can be embedded in these regions. These regions are selected based on human skin tone detection. Adaptive steganography are not an easy target for attacks especially when the hidden message is small (Chang et al., 2008). Embedding capacity and outstanding imperceptibility for the stego-images are successfully provided by the tri-way pixel value differencing method (Chang et al., 2008). The maximum PSNR that this scheme can attain is 38.89dB with approximate embedding capacity of 2 bpp. The result analysis shows that the proposed method achieves greater PSNR and better embedding capacity as compare to the scheme proposed by Chang et al. (2008).

Babu et al. (2008) proposed steganographic model authentication of secret information in image steganography that can be used to verify the integrity of the secret message from the stego-image. The payload in this method is transformed into spatial domain using discrete wavelet transform. The permutation of DWT coefficients are then embedded in the spatial domain of the cover image. This permutation is done with the verification code. DWT coefficients are used to generate the verification code. Thus the method can verify each row that has been modified by attacker.

Moon et al. (2007) proposed a fixed 4LSB method to embed an acceptable amount of data. It can easily be implemented and the degradation in the resulting image is not visually recognizable. However, the fundamental drawback of this scheme is that the encoded message can be easily recovered and even altered by third party. Lie et al. (1999) proposed an adaptive method of variable length bit substitution instead of fixed length to adjust the hiding capacity. Even though these methods (Chang et al., 2008; Babu et al., 2008; Moon et al., 2007; Lie et al., 1999) increase the embedding capacity as well as level of security, the visual distortion introduced is a cause of concern. The proposed scheme enhances the embedding capacity while reduces the visual distortion introduced.

The use of Exclusive-OR (XOR) operation and a binary-to-gray code conversion is proposed by Baekl et al. (2010). There are meaning patterns of the resultant Exclusive-OR operation and its relationship with the binary and gray code. This scheme requires that the N carrier images are shared amongst sender and receiver through a secure channel. Requirement of secure channel nullifies significance of steganography. So some methods are required so that the need of secured channel is nullified. The proposed scheme implements image steganography in such a way that the need for secure sharing of cover images is completely eliminated.

In (Lee et al., 2008) the author utilized a block-based lossless data embedding algorithm where the quantity of the hidden information each block bears is variable. The payload of each block depends on its cover image complexity which reduces the image distortion and increases the hiding capacity. Generally difference expansion schemes tend to damage the image quality seriously in the edge areas, so in (Lee et al., 2008) the author has chosen smoother areas to conceal more secret bits. The embedding capacity of this scheme is at most 1 bpp. The proposed scheme tries to achieve higher embedding capacity without violating the peak signal to noise ratio (PSNR) constraint.

## 3  Issues in steganography

Any steganography algorithm should consider two fundamental issues steganographic security measure and steganalysis. Steganographic security measure that any steganographic system incorporates has to be well defined. The system should satisfy any specific criteria applicable for steganographic security. Steganalysis deals with various analysis techniques employed on any algorithm to categorize the vulnerabilities associated with the algorithm.

### *3.1 Steganographic security measure*

There are different steganographic security measures as specified in (Cachin, 1998; Zollner et al., 1998). If one can distinguish between cover-image and stego-image, assuming one has unlimited computing power then the system is vulnerable to attacks. Let $P_C$ denote the probability distribution of cover-image and $P_S$ denote the probability distribution of stego-image. Cachin (1998) defines a steganographic algorithm to be $\varepsilon$-secure ($\varepsilon \leq 0$) if the relative entropy between the cover-object and the stego object probability distributions ($P_C$ and $P_S$, respectively) is at most $\varepsilon$.

$$D(P_C \parallel P_S) = \int P_C \cdot \log \frac{P_C}{P_S} \leq \varepsilon \qquad (1)$$

From this equation we note that detectability D(.) increases with the ratio $P_C/P_S$ which in turn means that the probability of steganalysis detection will also increase. A steganographic technique is said to be perfectly secure if $\varepsilon = 0$ (i.e. $P_C = P_S$). In this case the probability distributions of the cover and stego-objects are indistinguishable. Perfectly secure steganography algorithms are known to exist. In our proposed algorithm, we intend to achieve perfectly secure steganography with higher embedding capacity. The detectability function is more suitable for analyzing image steganography schemes where the embedding capacity is very low. More appropriate measure for visual distortion in image steganography with high embedding capacity is Peak Signal to Noise Ratio (PSNR).

### *3.2 Steganalysis*

There are two approaches to the problem of steganalysis; one is to come up with a steganalysis method specific to a particular steganographic algorithm. The other is developing techniques which are independent of the steganographic algorithm to be analyzed. Each of the two approaches has its own advantages and disadvantages. A steganalysis technique specific to an embedding method would give very good results when tested only on that embedding method, and might fail on all other steganographic algorithms. On the other hand, a steganalysis method which is independent of the embedding algorithm might perform less accurately overall but still provide acceptable results on new embedding algorithms. Steganalysis algorithms in essence are called successful if they can detect the presence of a message. The message itself does not have to be decoded. Indeed, the latter can be very hard if the message is encrypted using strong cryptography. However, recently there have been methods proposed in the literature which in addition to detecting the presence of a message are also able to estimate the size of the embedded message with great accuracy. We consider these aspects to be extraneous and only focus on the ability to detect the presence of a message. In our proposed algorithm we intend to foil most of the steganalysis methodologies if not all.

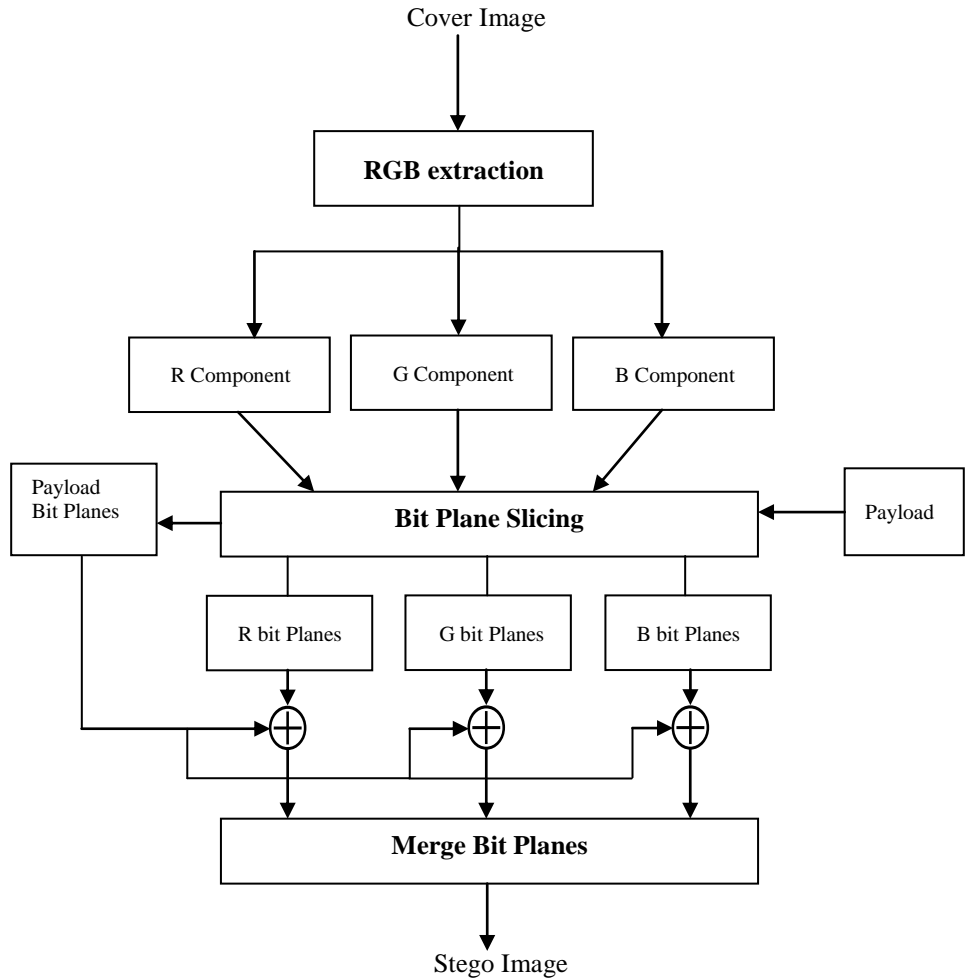

Figure 2: Bit plane X-ORing algorithm

# 4 Proposed Scheme

The higher embedding capacity is the focal point of the proposed scheme. The proposed scheme also eliminates the need of the cover image to recover the payload. Stego image is used to recover the payload. It reduces the need for a secure channel to share the cover images before the start of any secret communication. Figure 2 illustrates the operational flow of the proposed approach. The proposed algorithm has three phases. First phase includes bit plane slicing of the cover image as well as the payload into eight bit planes. This algorithm takes any RGB color image as a cover image, so RGB components are separated before the bit plane slicing. Second phase performs X-OR operation amongst some of the mid level bit planes of the cover image and high level bit planes of the payload and the result is injected into the remaining lower level bit planes of the cover image. The third phase recovers the payload from the stego image by X-ORing the mid level bit planes of the stego image with low level bit planes of the stego image. The resulting bit planes are the bit plane of the grayscale payload. First higher order two or three bit planes are considered as high level bit planes, next two or three higher bit planes are considered to be mid level bit planes.

A module Bit Plane Slicing extracts the bit planes of different component of the cover image and the bit planes of the payload. The module bit plane slicing has four sub modules R component bit plane slicing, G component bit plane slicing, B component bit plane slicing, and payload bit plane slicing.

## 4.1 First phase

Different components of the RGB cover image are separated in this phase. The major task of bit plane extraction of cover as well as payload image is done for further processing. This phase is operational at transmission end of any communication.

Let C be the RGB cover image. The R, G, and B components of the cover image C are separated. Let us denote these components as $C_R$, $C_G$ and $C_B$ respectively. Each component is sliced into eight bit planes.

$$(C_R | C_G | C_B) = \text{split\_comp}(C) \tag{2}$$

$$(C_R(1)|C_R(2)|C_R(3)|.....|C_R(7)|C_R(8)) = \text{dec2binp}(C_R, 8) \tag{3}$$

$$(C_G(1)|C_G(2)|C_G(3)|.....|C_G(7)|C_G(8)) = \text{dec2binp}(C_G, 8) \tag{4}$$

$$(C_B(1)|C_B(2)|C_B(3)|.....|C_B(7)|C_B(8)) = \text{dec2binp}(C_B, 8) \tag{5}$$

The function *split_comp()* separates three components of an RGB image. Equations (2)-(4) slices each component into eight bit planes, where $C_R(1)$, $C_G(1)$ and $C_B(1)$ are the highest level bit planes of $C_R$, $C_G$, and $C_B$ respectively and $C_R(8)$, $C_G(8)$ and $C_B(8)$ are the lowest level bit planes of $C_R$, $C_G$, and $C_B$ respectively. Similarly, payload P is sliced as in equation (5).

$$(P(1)|P(2)|P(3)|P(4)|P(5)|P(6)|P(7)|P(8)) = \text{dec2binp}(P, 8) \tag{6}$$

The function *dec2binp()* converts any plane in decimal into it's binary equivalent planes and returns an array of planes of size as specified in the second parameter of the function.

*4.2 Second phase*

This phase performs Bit plane X-ORing on different bit planes of cover and payload images. First three bit planes of payload P; P(1), P(2), and P(3) are X-ORed with mid level three bit planes of $C_R$ and the result is stored in the low level three bit planes of $C_R$. The basic idea behind X-ORing the mid level bit planes of cover image with the high level bit planes of the payload is that some of the characteristics of mid level bit planes remain intact in the resulting bit planes due to the nature of the X-OR operation. When we insert these planes in the lower level bit planes of the cover image, we are essentially transforming the bit patterns of the lower level bit planes into the bit patterns of the mid level bit planes of the cover image. This diminishes the expected distortion in the resulting stego image.

$$C_{RM}(6) = XOR(C_R(5), P(1)) \tag{7}$$
$$C_{RM}(7) = XOR(C_R(4), P(2)) \tag{8}$$
$$C_{RM}(8) = XOR(C_R(3), P(3)) \tag{9}$$

Similarly, P(4), P(5), and P(6) are X-ORed with mid level three bit planes of $C_G$ and the result is stored in the low level three bit planes of $C_G$.

$$C_{GM}(6) = XOR(C_G(5), P(4)) \tag{10}$$
$$C_{GM}(7) = XOR(C_G(4), P(5)) \tag{11}$$
$$C_{GM}(8) = XOR(C_G(3), P(6)) \tag{12}$$

Last two bit planes of the payload are X-ORed with mid level two bit planes of $C_B$ and the result is stored in the low level two bit planes of $C_B$.

$$C_{BM}(7) = XOR(C_B(6), P(7)) \tag{13}$$
$$C_{BM}(8) = XOR(C_B(5), P(8)) \tag{14}$$

Red, green and blue components of the stego image are reconstructed using the high level bit planes and modified mid level bit planes of each component of the cover image.

$$C_{RM} = bin2decp(C_R(1)|C_R(2)|C_R(3)|.....|C_{RM}(7)|C_{RM}(8)) \tag{15}$$
$$C_{GM} = bin2decp(C_G(1)|C_G(2)|C_G(3)|.....|C_{GM}(7)|C_{GM}(8)) \tag{16}$$
$$C_{BM} = bin2decp(C_B(1)|C_B(2)|C_B(3)|.....|C_{BM}(7)|C_{BM}(8)) \tag{17}$$

The function *bin2decp()* converts an array of binary planes into a decimal plane. Stego image S is formed by merging these three modified components.

$$S = Merge\_Comp(C_{RM}, C_{GM}, C_{BM}) \tag{18}$$

*Merge_Comp()* merges three components to form a RGB image. So, S is the stego image obtained after phase three.

*4.3 Third phase*

This phase is the recovery phase. Payload is recovered from the stego image at the receiving end. Recovery process is just the reverse of phase two. Stego image is split into the equivalent red, green and blue component.

$$(C_{RM}|C_{GM}|C_{BM}) = \text{split\_comp}(S) \tag{19}$$

The function *split_comp()* separates three components of an RGB image. Now, individual color components are sliced into bit planes.

$$(C_{RM}(1)|C_{RM}(2)|.....|C_{RM}(7)|C_{RM}(8)) = \text{dec2binp}(C_{RM}, 8) \tag{20}$$

$$(C_{GM}(1)|C_{GM}(2)|.....|C_{GM}(7)|C_{GM}(8)) = \text{dec2binp}(C_{GM}, 8) \tag{21}$$

$$(C_{BM}(1)|C_{BM}(2)|.....|C_{BM}(7)|C_{BM}(8)) = \text{dec2binp}(C_{BM}, 8) \tag{22}$$

The payload bit planes are recovered by X-ORing the respective bit planes of each component.

$$RP(1) = \text{XOR}(C_{RM}(5), C_{RM}(6)) \tag{23}$$

$$RP(2) = \text{XOR}(C_{RM}(4), C_{RM}(7)) \tag{24}$$

$$RP(3) = \text{XOR}(C_{RM}(3), C_{RM}(8)) \tag{25}$$

$$RP(4) = \text{XOR}(C_{GM}(5), C_{GM}(6)) \tag{26}$$

$$RP(5) = \text{XOR}(C_{GM}(4), C_{GM}(7)) \tag{27}$$

$$RP(6) = \text{XOR}(C_{GM}(3), C_{GM}(8)) \tag{28}$$

$$RP(7) = \text{XOR}(C_{BM}(6), C_{BM}(7)) \tag{29}$$

$$RP(8) = \text{XOR}(C_{BM}(5), C_{BM}(8)) \tag{30}$$

These recovered bit planes are combined to recover the payload.

$$RP = \text{bin2decp}(RP(1)|RP(2)|RP(3).....|RP(7)|RP(8)) \tag{31}$$

Equation (31) gives the recovered payload. This payload is obtained from the stego image and this method does not require any cover image to be shared.

## 5 Experimental results

To check the performance of the proposed method we embedded the same payload in six different cover images Lena, Barbara, Boat, Greens, Pepper, and Baboon of different complexity. All the cover images were selected from standard image set used in other state-of-art. The cover images and the payload are shown in Figure 3(a)-(f) and Figure 3(g) respectively. In each instance the size of the cover image $(M \times N)$ and the size of the payload $(m \times n)$ satisfied the following constraints $m \leq M \ \& \ n \leq N$.

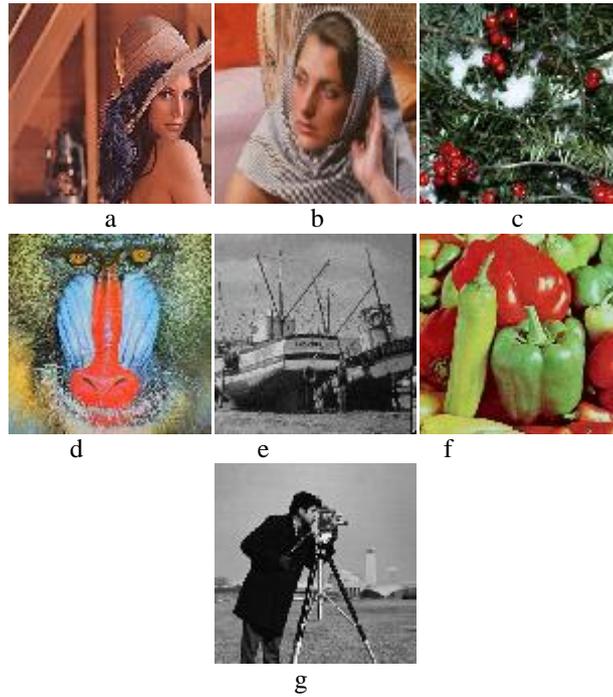

Figure 3: Test images (a)-(f) and Payload (g). (a) Lena; (b) Barbara; (c) Greens; (d) Baboon; (e) Boat; (f) Pepper; (g) Cameraman.

By taking a cover image of size slightly greater than the size of payload the noise in the sego image can be reduced further. The secret bit planes are produced by performing X-OR operation amongst the bit planes of the cover image and the bit planes of the payload. The overall payload is same and independent of the complexity of the cover image. With our scheme 8 bits of the secret data are inserted in each pixel of the cover image, meaning there by 8 bit pixel information of the payload is embedded in 24 bit pixel information of the cover image. The major problem in increasing the embedding capacity is the rise in visual distortion in the stego image. It is generally known that the distortion of the stego image is hard to detect by the human eyes as long as the PSNR value is greater than or equal to 30 dB (Lee et al., 2008). The amount of data inserted in each instance of the experiments is kept constant to analyze the distortion introduced in the stego image, while the cover image in each instance is changed. Even though, the distortion is visually unrecognizable the introduction of noise is inevitable.

The fundamental method used to determine the noise in stego image is peak-signal-to-noise-ratio (PSNR). Efficiency of any image steganography algorithm depends on hiding capacity and embedding efficiency. So, we consider both aspects to analyze the results. PSNR is an objective measure for subjective evaluation of degree of similarity between an original image and a stego image (Baekl et al., 2010). PSNR is defined as;

$$PSNR = 10 \times \log_{10}\left(\frac{I_{max}^2}{MSE}\right)(dB) \qquad (32)$$

where $I_{max} = 255$, maximum gray level for any grayscale image, and the mean squared error MSE (Moon et al., 2007) is defined to be

$$MSE = \frac{1}{MN}\sum_{i=1}^{M}\sum_{j=1}^{N}(|C(i,j) - S(i,j)|)^2 \qquad (33)$$

where M and N represent the number of horizontal and vertical pixels of the images respectively. We are using a RGB image as cover image, so the noise introduced in each component of the stego image has to be evaluated. The PSNR of the stego image is defined as the average of the PSNR calculated for different components of the stego image.

Let $C_R(i, j)$ be the pixel intensity of R component of the cover image and $S_R(i, j)$ be the pixel intensity of R component of the stego image. Similarly for green and blue components we have $C_G(i, j)$, $C_B(i, j)$ and $S_G(i, j), S_B(i, j)$.

MSE for R, G, and B components are calculated using following equations:

$$MSE_R = \frac{1}{MN}\sum_{i=1}^{M}\sum_{j=1}^{N}(|C_R(i,j) - S_R(i,j)|)^2 \qquad (34)$$

$$MSE_G = \frac{1}{MN}\sum_{i=1}^{M}\sum_{j=1}^{N}(|C_G(i,j) - S_G(i,j)|)^2 \qquad (35)$$

$$MSE_B = \frac{1}{MN}\sum_{i=1}^{M}\sum_{j=1}^{N}(|C_B(i,j) - S_B(i,j)|)^2 \qquad (36)$$

PSNR for R, G, and B components are calculated using following equations:

$$PSNR_R = 10 \times \log_{10}\left(\frac{I_{max}^2}{MSE_R}\right)(dB) \qquad (37)$$

$$PSNR_G = 10 \times \log_{10}\left(\frac{I_{max}^2}{MSE_G}\right)(dB) \qquad (38)$$

$$PSNR_B = 10 \times \log_{10}\left(\frac{I_{max}^2}{MSE_B}\right)(dB) \qquad (39)$$

The PSNR is the average of $PSNR_R$, $PSNR_G$, and $PSNR_B$.

$$PSNR = \frac{PSNR_R + PSNR_G + PSNR_B}{3} \qquad (40)$$

Figure 4(a)-(l) shows the cover and the resulting stego images. The image "Lena" is a cover image with high complexity even then we keep the size of payload same. Each instance of the experiment has the same payload (Cameraman). Even though eight pixel information of the payload is distributed over different bit planes of the cover image, the visual distortion is almost unrecognizable.

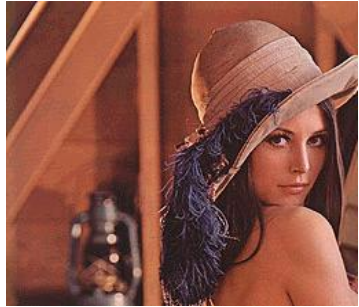
a

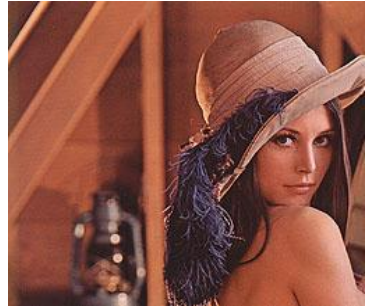
g

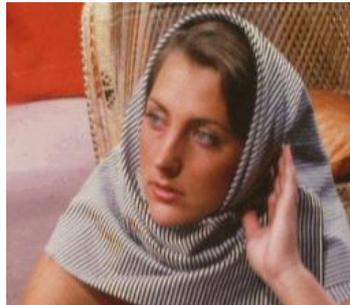
b

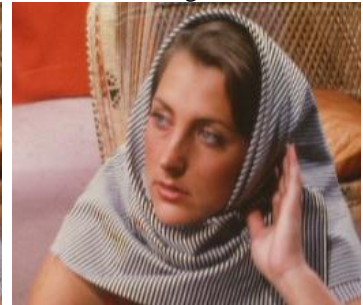
h

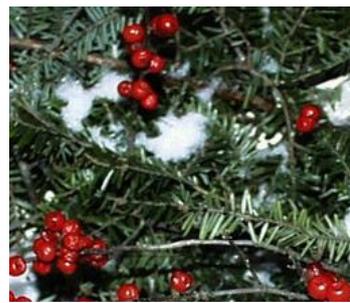
c

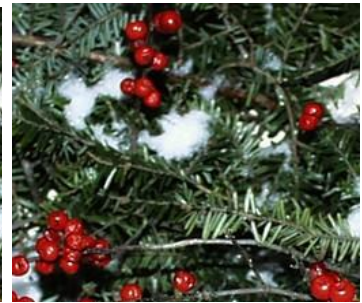
i

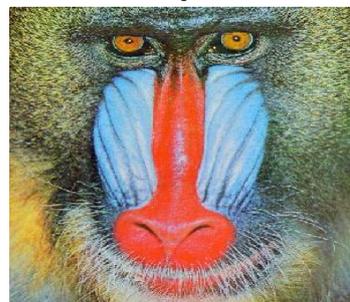
d

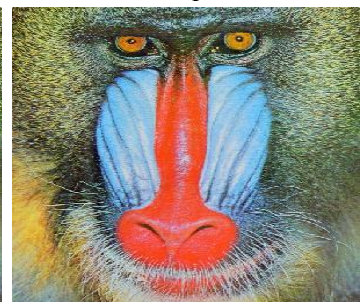
j

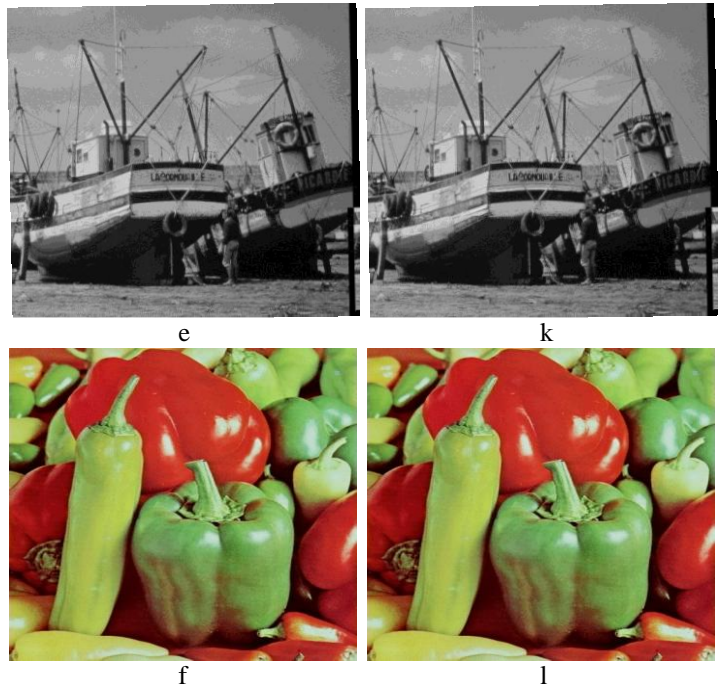

|   |   |
|---|---|
| e | k |
| f | l |

Figure 4: Cover images (a)-(f) and stego images (g)-(l)

Change in cover image does not affect the visual features of the payload after extraction. Visual distortion in the stego image is minimal, if we change the cover image for the same payload. Complexity of the image "Barbara" is different in many respects even then the distortion in the stego image for the same payload (Cameraman) is negligible. The fundamental requirement of any image steganography algorithm is minimum visual distortion in the resulting stego image and it should be same for different cover images. The results illustrate that the proposed algorithm conforms to these requirements.

**Table 1** Mean Square Error for different components of the cover image and Payload

| Images | Mean Square Error (decimal) | | |
|---|---|---|---|
|  | Red | Green | Blue |
| Barbara | 8970 | 8190 | 9330 |
| Lena | 6070 | 6130 | 7210 |
| Green | 10100 | 9480 | 11300 |
| Baboon | 8350 | 6420 | 8740 |
| Boat | 6164 | 6164 | 6164 |
| Pepper | 6522 | 10525 | 9092 |

**Table 2** Comparison results (Lin et al., 2004), (Lee et al., 2008), and proposed scheme

| Images | Embedded Data | PSNR(dB) (Lin et al., 2004) | PSNR(dB) (Lee et al., 2008) | Average PSNR(dB) Proposed Scheme |
|---|---|---|---|---|
| Barbara | 4,60,800 bits | 38.12 | 34.74 | 40.08 |
| Lena |  | 38.52 | 34.32 | 39.94 |
| Greens |  | 38.38 | 34.27 | 40 |
| Baboon |  | 38.26 | 34.84 | 40.04 |
| Boat |  | 38.40 | 34.41 | 40.07 |
| Pepper |  | 38.45 | 34.24 | 40.04 |

**Table 3** Comparison of PSNR for different cover images and different size payload using proposed scheme

| Payload (Bits) | Barbara PSNR(dB) | Lena PSNR(dB) | Greens PSNR(dB) | Baboon PSNR(dB) | Boat PSNR(dB) | Pepper PSNR(dB) |
|---|---|---|---|---|---|---|
| 20,000 | 40.05 | 39.97 | 39.94 | 39.89 | 40.07 | 40.05 |
| 80,000 | 40.09 | 39.95 | 39.93 | 40 | 39.99 | 40.02 |
| 3,20,000 | 40.05 | 39.96 | 39.99 | 40.04 | 40.01 | 40.06 |
| 4,60,800 | 40.08 | 39.94 | 40 | 40.04 | 40.07 | 40.04 |

**Table 4** Comparison of Embedding Capacity and PSNR for state of art and the proposed scheme

| Steganography Schemes | Maximum Embedding Capacity (bpp) | Maximum PSNR (dB) |
|---|---|---|
| (Lin et al., 2004) | 1 | 38.52 |
| (Chang et al., 2008) | 1 | 38.89 |
| (Babu et al., 2008) | 0.99 | 50.13 |
| 4 LSB | 4 | 36.67 |
| (Lie et al., 1999) | 0.45 | 40 |
| (Baekl et al., 2010) | 1 | 58.42 |
| (Lee et al., 2008) | 1 | 34.84 |
| Proposed Scheme | 8 | 40 |

Mean Square Error (MSE) of different components of cover images are shown in Table 1. Table 1 illustrates that the MSE of blue component is greatest in most of the cases. So in the proposed scheme significant noise reduction in stego image has been achieved by embedding minimum number of bit planes in blue component of the cover image.

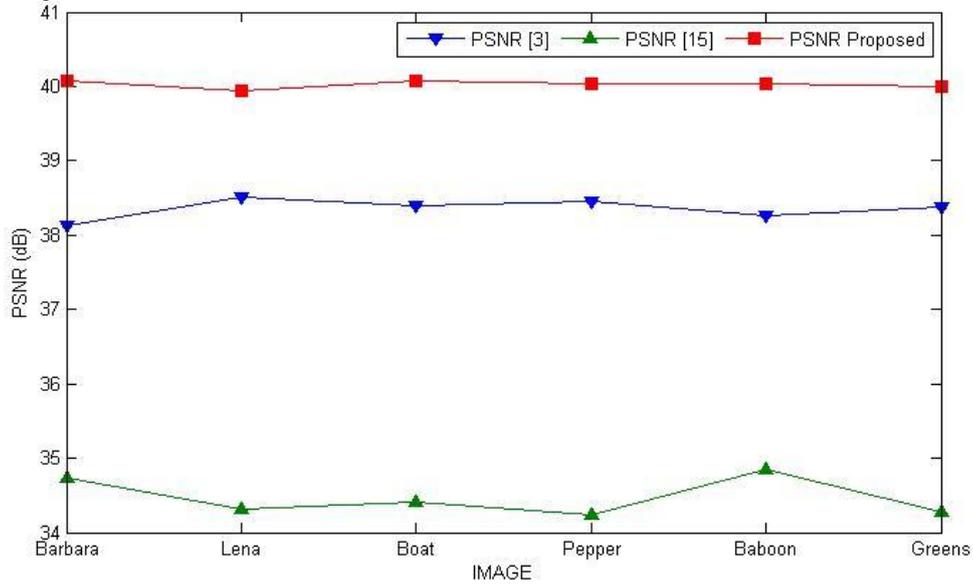

Figure 5: Comparison results (Lin et al., 2004), (Lee et al., 2008), and the proposed scheme

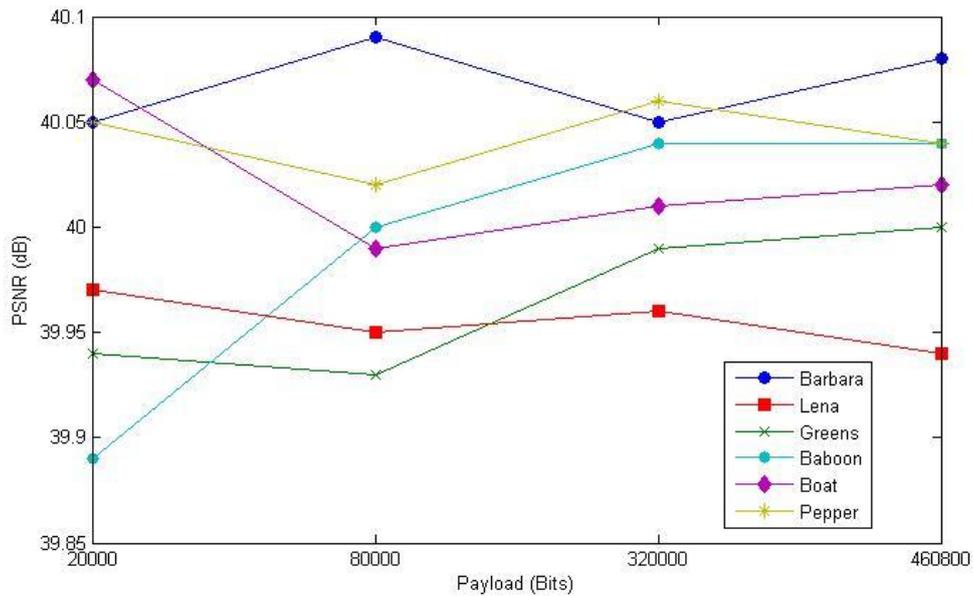

Figure 6: Comparison of PSNR for different cover images and different payload using proposed algorithm.

Results shown in Table 2 illustrate comparison of proposed scheme with (Lin et al., 2004) and (Lee et al., 2008). The proposed scheme achieves approximately same PSNR (dB) for different cover images, for the same payload. Hence, the choice of cover image is not a critical consideration for the proposed algorithm. A single image is enough to carry the entire payload and one need not consider a series of images for data embedding. For different sized payload, the change in PSNR for different cover images is shown in Table 3. Results shown in Table 3 clearly illustrate the fact that for a given cover image the PSNR is stable for different sized payloads. Maximum embedding capacity and the maximum PSNR achieved in state of art and proposed scheme has been depicted in Table 4. The proposed method attains higher embedding capacity while maintaining the desired PSNR. The proposed algorithm is efficient enough in attaining higher and stable PSNR as compared to (Lin et al., 2004) and (Lee et al., 2008) as shown in Figure 5.The change in PSNR for different payload is shown in Figure 6. For increasing payload the proposed algorithm confer minimum change in PSNR. Even if the size of the payload is increased, the PSNR value never goes below the minimum required value of 30dB.

## 6 Conclusion

This paper proposes an algorithm that exploits the multiple dimensions of a RGB image to embed a secret image. The method not only achieves high bit per pixel embedding but also manages reduced MSE in the resulting stego-image. Bit Plane X-ORing is used to modify the bit planes of the different components of the cover image. The resulting stego image is of the same size as the size of the cover image, which nullifies the detection of hidden information in the carrier. Recovery method involves X-ORing of certain bit plains of the stego image. The proposed method attains higher PSNR and embedding capacity. In this scheme the need for shared cover images for recovery of the payload is taken out of the picture. Further improvements are possible on this method if any other correlation could be found amongst the R, G, and B components of the cover-image and the payload. Our future work will involve working with color images as payload. A method also needs to be developed for grayscale payload images whose pixel depth is more than 8 pixels, as is the case in technical uses like medical imaging or remote sensing.